\documentclass[aps,preprint,showpacs,showkeys]{revtex4}
\usepackage[pdftex]{graphicx}
\usepackage{amssymb}
\usepackage{bm}
\begin{document}
\setcounter{page}{1}
\title[]{Hillery-Type Squeezing in Fan-States}
\author{Minh Duc \surname{Truong}}
\affiliation{Abdus Salam International Center for Theoretical Physics, Strada Costiera 11, 34100 Trieste, Italy}
\author{Ba An \surname{Nguyen}}
\email{nbaan@kias.re.kr}
\thanks{Fax: +82-2-958-3786; Corresponding author}
\affiliation{School of Computational Sciences, Korea Institute for Advanced Study, 207-43 Cheongryangni 2-dong, Dongdaemun-gu, Seoul 130-722, Korea}
\date[]{}

\begin{abstract}
We study the Hillery-type, i.e. $N$-th power, amplitude squeezing in the fan-state 
$\left| \xi ;2k,f\right>_{F}$ characterized by $\xi \in \mathcal{C}$, $k=1,2,3,...$ 
and $f$ a nonlinear operator-valued function. We show that for a given $k$ 
there exists a critical $\xi_c$ such that for $0<|\xi|\leq|\xi_c|$ squeezing occurs 
simultaneously in $2k$ directions for the powers $N$ which are a multiple of $2k$. 
This result does not depend on the concrete form of $f$, i.e. it holds true for both 
$f\equiv 1$ and $f\neq 1$. However, for $f\neq 1$, which is realized here in the 
ion trap context, the squeezing directions as well as the magnitude of $\xi$ 
can be controlled by adjusting the system driving parameters.
\end{abstract}



\maketitle

\section{INTRODUCTION}

The squeezed state is a nonclassical state, which has been discovered for a long
time (see, e.g., a recent review \cite{r1} and references therein). Besides
recognized potential applications in detecting extremely weak forces, squeezed states
have recently attracted much more attention in the newly emerging field of
quantum information and quantum computation. By means of squeezed states schemes for teleportation
of continuous quantum variables \cite{t1,t2,t3}, quantum cryptography \cite
{cry1,cry2}, entanglement distribution \cite{ent}, etc. have been proposed
and some of them have been realized. The original squeezed state \cite{r8} has been generalized
to different types of higher-order. One type of higher-order amplitude
squeezing is suggested by Hong and Mandel \cite{r9}. Another qualitatively
different type is introduced by Hillery \cite{r10} which is then developed
further by several authors (see, e.g., \cite{r11,r12,r13}). The
Hong-Mandel type higher-order amplitude squeezing has been investigated in the
fan-state $\left| \xi ;2k,f\right\rangle _{F}$ which is introduced in \cite
{r14} as a linear superposition of $2k$ $2k$-quantum nonlinear coherent
states $\left| \xi _{q};2k,f\right\rangle $ $(q=0,1,...,2k-1)$ \cite{r15} in
a phase-locked manner. Since there are states that exhibit Hong-Mandel type squeezing 
but not Hillery-type one and vice versa \cite{r10}, we would like to examine 
in this paper whether the Hillery-type $N$-th power
amplitude squeezing is possible and, if it is,  how it behaves in the same fan-state. 
The normalized fan-state is defined as 
\begin{equation}
|\xi ;2k,f\rangle _{F}=\frac{1}{\sqrt{D_{k}(|\xi|^2)}}\sum_{q=0}^{2k-1}\left| \xi
_{q};2k,f\right\rangle ,  \label{fan}
\end{equation}
where $k=1,2,3,...;$ $\xi _{q}=\xi \exp (\frac{i\pi q}{2k}),$ $\xi \in 
\mathcal{C},$ $f$ is in general an arbitrary real nonlinear operator-valued
function (in particular, it may be that $f\equiv 1$) 
of $\hat{n}=a^{\mathbf{\dagger }}a$ with $a$ $(a^{\mathbf{\dagger }%
})$ the bosonic annihilation (creation) operator, 
\begin{equation}
D_{k}(|\xi |^{2})=\sum_{m=0}^{\infty }\frac{|\xi |^{4km}|J_{k}(m)|^{2}%
}{(2km)![f(2km)(!)^{2k}]^{2}},  \label{e1}
\end{equation}
with 
\begin{equation}
J_{k}(m)=\sum_{n=0}^{2k-1}\text{e}^{i\pi nm}
\end{equation}
 and the $2k$-quantum
nonlinear coherent state is given by 
\begin{equation}
\left| \xi _{q};2k,f\right\rangle =\sum_{n=0}^{\infty }\frac{\xi _{q}^{2kn}}{%
\sqrt{(2kn)!}f(2kn)(!)^{2k}}\left| 2kn\right\rangle   \label{e2}
\end{equation}
with $\left| 2kn\right\rangle $ a Fock state. The state $\left| \xi
_{q};2k,f\right\rangle $ is a sub-state of the multi-quantum nonlinear
coherent state which is by definition \cite{r15,r16,r17} the eigenstate of the
nonboson operator $a^{2k}f(\hat{n}).$ The notation $(!)^{2k}$ appearing in 
Eqs. (\ref{e1}) and (\ref{e2}) is understood as follows 
\[
f(p)(!)^{2k}=\left\{ 
\begin{array}{ll}
f(p)f(p-2k)f(p-4k)...f(q) & \text{if}\hspace{0.15cm}p\geq 2k;0\leq q<2k \\ 
1 & \text{if}\hspace{0.15cm}0\leq p<2k
\end{array}
\right. .
\]
The Hillery-type $N$-th power amplitude squeezing is associated with the operator $%
Q_{N}(\varphi )$ of the form 
\begin{equation}
Q_{N}(\varphi )=\frac{1}{2}(a^{N}e^{-iN\varphi }+a^{\mathbf{\dagger }%
N}e^{iN\varphi })
\end{equation}
with $\varphi $ an angle determining the direction of $\langle Q_{N}(\varphi
)\rangle $ in the complex plane. According to \cite{r12,r13}, a state $%
\left| ...\right\rangle $ is said to be $N$-th power amplitude squeezed
along the direction $\varphi $ if there exists a value of $\varphi $ such
that the variance of $Q_{N}(\varphi )$ satisfies the following inequality 
\begin{equation}
\langle (\Delta Q_{N}(\varphi ))^{2}\rangle <\frac{1}{4}\langle F_{N}\rangle
\label{r3}
\end{equation}
where
\begin{equation}
\langle F_{N}\rangle  =\langle [a^{N},a^{\mathbf{\dagger }N}]\rangle .
\label{rr3}
\end{equation}
It is easy to express the variance as 
\begin{equation}
\langle (\Delta Q_{N}(\varphi ))^{2}\rangle =\frac{1}{4}\langle F_{N}\rangle
+\langle :(\Delta Q_{N}(\varphi ))^{2}:\rangle   \label{r4}
\end{equation}
where 
\begin{equation}
\langle :(\Delta Q_{N}(\varphi ))^{2}:\rangle =\frac{1}{2}\{\langle a^{%
\mathbf{\dagger }N}a^{N}\rangle +\Re {[}e^{-2iN\varphi }{\langle }a^{2N}{%
\rangle ]}-2\left( \Re [e^{-iN\varphi }{\langle }a^{N}{\rangle ]}\right)
^{2}\}  \label{r5}
\end{equation}
with $:...:$ denotes a normal ordering of the operators. The $\langle F_{N}\rangle$ 
can also be normally ordered and its explicit form reads  \cite{r13}
\begin{equation}
\langle F_{N}\rangle =\sum_{q=1}^{N}\frac{N!N^{(q)}}{(N-q)!q!}\langle (a^{%
\mathbf{\dagger }})^{N-q}a^{N-q}\rangle   \label{r6}
\end{equation}
with $N^{(q)}=N(N-1)...(N-q+1)$. Combining the inequality (\ref{r3}) and Eq. (\ref{r4}) we
recognize that the state becomes squeezed whenever $\langle :(\Delta
Q_{N}(\varphi ))^{2}:\rangle $ gets negative. For convenience, we define a measure
of squeezing degree by a quantity $S$ scaled as 
\begin{equation}
S=\frac{4\langle :(\Delta Q_{N}(\varphi ))^{2}:\rangle }{\langle
F_{N}\rangle }.  \label{S}
\end{equation}
Clearly that for a squeezed state $-1\leq S<0$ and the ideal squeezing
corresponds to $S=-1.$ Without any loss of generality we choose the real axis to be along the direction of $\xi $
in the phase space. This allows us to treat $\xi $ as a real number. In the
fan-state, we have 
\begin{equation}
\langle a^{\mathbf{\dagger }l}a^{m}\rangle _{k}=\frac{\xi ^{l-m}}{%
D_{k}(\xi ^{2})}I(\frac{l-m}{2k})\sum_{n=0}^{\infty }\frac{\theta (2kn-m)\xi
^{4kn}J_{k}(n+\frac{l-m}{2k})J_{k}(n)}{%
(2kn-m)!f(2kn)(!)^{2k}f(2kn+l-m)(!)^{2k}}  \label{alm}
\end{equation}
where $\langle ...\rangle _{k}\equiv \,_{F}\langle \xi ;2k,f\mid ...\mid \xi
;2k,f\rangle _{F}$. The function $I(x)$ equals unity (zero) if $x$ is an
integer (non-integer) and $\theta (x)$ is the step function: $\theta(x)=1$ $(0)$ for $x\geq 0$ $(x<0)$. 
Putting Eqs. (\ref{r5}) and  (\ref{r6}) into Eq. (\ref{S}) and making a simple trigonometric
transformation, we arrive explicitly at
\begin{equation}
S=\frac{2\{\langle a^{\mathbf{\dagger }N}a^{N}\rangle _{k}-\langle
a^{N}\rangle _{k}^{2}+\cos (2N\varphi )(\langle a^{2N}\rangle _{k}-\langle
a^{N}\rangle _{k}^{2})\}}{\sum_{q=1}^{N}\frac{N!N^{(q)}}{(N-q)!q!}\langle
(a^{\mathbf{\dagger }})^{N-q}a^{N-q}\rangle _{k}}  \label{SS}
\end{equation}
where $\langle a^{N}\rangle _{k},$ $\langle a^{2N}\rangle _{k},$ $\langle a^{%
\mathbf{\dagger }N}a^{N}\rangle _{k}$ and $\langle (a^{\mathbf{\dagger }%
})^{N-q}a^{N-q}\rangle _{k}$ are determined by (\ref{alm}) if $\{l,m\}$ is
set to be $\{0,N\},$ $\{0,2N\},$ $\{N,N\}$ and $\{N-q,N-q\}.$

In what follows we distinguish two cases: $f\equiv 1$ which corresponds to
light field and $f\neq 1$ which may be associated with the vibrational
motion of the center-of-mass of a trapped ion.

\section{CASE $f\equiv 1$}

A QED-based scheme to generate the fan-state with $f\equiv 1$ which
concerns radiation field in a cavity has been proposed in \cite{f1}. For $%
f\equiv 1$ the expectation values $\langle a^{N}\rangle _{k}$ and $\langle
a^{2N}\rangle _{k}$ simplify to 
\begin{equation}
\langle a^{N}\rangle _{k}=\frac{\xi ^{-N}}{D_{k}(\xi ^{2})}I(-\frac{N}{2k}%
)\sum_{n=0}^{\infty }\frac{\theta (2kn-N)\xi ^{4kn}J_{k}(n-\frac{N}{2k}%
)J_{k}(n)}{(2kn-N)!},  \label{aN}
\end{equation}
\begin{equation}
\langle a^{2N}\rangle _{k}=\frac{\xi ^{-2N}}{D_{k}(\xi ^{2})}I(-\frac{N}{k}%
)\sum_{n=0}^{\infty }\frac{\theta (2kn-2N)\xi ^{4kn}J_{k}(n-\frac{N}{k}%
)J_{k}(n)}{(2kn-2N)!}.  \label{a2N}
\end{equation}
Using the property of the function $J_{k}(n),$%
\begin{equation}
J_{k}(n)J_{k}(n\pm n^{\prime })=\left\{ 
\begin{tabular}{ll}
$2k^{2}(1+(-1)^{n})$ & if $n^{\prime }$ is an even integer \\ 
$0$ & if $n^{\prime }$ is an odd integer
\end{tabular}
\right. ,  \label{J}
\end{equation}
it follows from Eqs. (\ref{aN}) and (\ref{a2N}) that, for a given $k,$ both $%
\langle a^{N}\rangle _{k}$ and $\langle a^{2N}\rangle _{k}$ vanish if $N\neq
2kp$ with $p=1,2,3,....$ On the other hand, 
\begin{equation}
\langle a^{\mathbf{\dagger }N}a^{N}\rangle _{k}=\frac{4k^{2}}{D_{k}(\xi ^{2})%
}\sum_{n=0}^{\infty }\frac{\theta (4kn-N)\xi ^{8kn}}{(4kn-N)!}  \label{aNN}
\end{equation}
is always positive. Hence, for $N\neq 2kp$ the $\varphi$-dependence in $S$ (see Eq. (\ref{SS})) 
is killed and we always obtain $S>0$, i.e. squeezing is impossible. 
As a consequence, squeezing turns out to be possible only for powers $N$ that are a multiple
of $2k,$ a necessary condition for the Hillery-type squeezing in the fan-state. 
 Yet, whether or not squeezing appears still depends on the value of 
$\xi .$ To see this let us analytically calculate $S$ for a couple of concrete
values of $k$ and $N.$

For $k=1$ and $N=2$ we have obtained explicitly 
\begin{equation}
S_{\varphi ,N=2}^{(k=1)}=\frac{\xi ^{4}[\cosh (\xi ^{2})-\cos (\xi
^{2})+D_{1}(\xi^2)\cos (4\varphi )]}{2\xi ^{2}(\sinh (\xi ^{2})-\sin (\xi
^{2}))+D_{1}(\xi^2)}
\end{equation}
with 
\begin{equation}
D_{1}(\xi^2)=2[\cos(\xi^2)+\cosh(\xi^2)]
\end{equation}
and squeezing occurs whenever 
\begin{equation}
\cos (4\varphi )<h(\xi ^{2})=\frac{\cos (\xi ^{2})-\cosh (\xi ^{2})}{%
D_{1}(\xi ^{2})}\leq 0.
\end{equation}
The function $h(\xi ^{2})$ equals zero at $\xi =0$ and decreases for
increasing $|\xi |$. There is no squeezing for $|\xi | > \xi _{c}=1.2533$
for which $h(\xi ^{2}) < -1$ and no $\varphi $ can be found to make $%
S_{\varphi ,N=2}^{(k=1)}$ negative. This defines the range of $\xi$ within 
which squeezing takes place: $0<|\xi| \leq |\xi_c|.$ Figure 1 is a 3D plot of $S_{\varphi
,N=2}^{(k=1)}$ as a function of $|\xi |$ and $\varphi $. Interestingly to
notice that a maximal squeezing occurs simultaneously along two directions
characterized by $\varphi _{1}=\pi /4$ and $\varphi _{2}=3\pi /4.$ The two
coexistent directions of squeezing can alternatively be seen by a polar plot of $S_{\varphi
,N=2}^{(k=1)}$ in Fig. 2 with $|\xi |=0.8$ which looks like an eight-winged
bow. The shorter wings correspond to squeezing, while the longer ones to stretching. 
To better understand Fig. 2 let us also show in Fig. 3 the variance 
$\left< \left( \Delta Q_N(\varphi) \right)^2\right>$ itself as a function of $\varphi $ 
for the same parameters as in Fig. 2, i.e. 
for $\{ k,N,|\xi|\}=\{ 1,2,0.8\}$. Taking into account Eqs. (\ref{r4}) and (\ref{S}) yields 
\begin{equation}
\left< \left( \Delta Q_N(\varphi) \right)^2\right>_k = 
\frac{\left< F_N \right>_k}{4}\left( 1+S^{(k)}_{\varphi ,N}\right)
\end{equation} 
In Fig. 3 the variance $\left< \left( \Delta Q_N(\varphi) \right)^2\right>_k$ 
is represented by the solid curve, while the dashed circle of 
radius $\left< F_N \right>_k/4$  determines the uncertainty region associated 
with the coherent state for which $S^{(k)}_{\varphi ,N}=0$ (Note, this corresponds 
to the center point in Fig. 2). It is clear that, along the directions 
$\varphi=\pi/4$ and $\varphi=3\pi/4,$  we have 
$\left< \left( \Delta Q_N(\varphi) \right)^2\right>_k<\left< F_N \right>_k/4$  signifying 
negativity of $S^{(k)}_{\varphi ,N},$ i.e. squeesing takes place. 
Along the directions $\varphi=0$ and $\varphi=\pi/2,$ however, the inequality 
 $\left< \left( \Delta Q_N(\varphi) \right)^2\right>_k > \left< F_N \right>_k/4$ holds.  
That implies $S^{(k)}_{\varphi ,N}>0,$ i.e. no squeesing occurs. Since the difference between 
the solid curve and the dashed circle along $\varphi=\pi/4$ (and $3\pi/4 $) is smaller than that 
along $\varphi=0$ (and $\pi/2$), the squeezing (stretching) gives rise to a shorter (longer) wing 
in Fig. 2. For $N=4,$ $6,$ $8,$ $...$ simultaneous squeezing along the same two directions
may also result with appropriate values of $\xi $ but the squeezing degree
is much lower than that in the case with $N=2.$

For $k=2$ and $N=4$ we have obtained explicitly 
\begin{equation}
S_{\varphi ,N=4}^{(k=2)}=\frac{\xi ^{8}}{4}\frac{\cosh (\xi ^{2})+\cos (\xi
^{2})-2\cosh (\frac{\xi ^{2}}{\sqrt{2}})\cos (\frac{\xi ^{2}}{\sqrt{2}}%
)+D_{2}(\xi ^{2})\cos (8\varphi )}{3\xi ^{4}[\cosh (\xi ^{2})-\cos (\xi
^{2})-2\sinh (\frac{\xi ^{2}}{\sqrt{2}})\sin (\frac{\xi ^{2}}{\sqrt{2}}%
)]+A(\xi ^{2})+2B(\xi ^{2})-D_{2}(\xi ^{2})},
\end{equation}
where 
\begin{equation}
D_{2}(\xi ^{2})=\cosh (\xi ^{2})+\cos (\xi ^{2})+2\cosh (\frac{\xi ^{2}}{%
\sqrt{2}})\cos (\frac{\xi ^{2}}{\sqrt{2}}),
\end{equation}
\begin{equation}
A(\xi ^{2})=2\xi ^{6}(\sinh (\xi ^{2})+\sin (\xi ^{2})-\sqrt{2}[\sinh (\frac{%
\xi ^{2}}{\sqrt{2}})\cos (\frac{\xi ^{2}}{\sqrt{2}})+\sin (\frac{\xi ^{2}}{%
\sqrt{2}})\cosh (\frac{\xi ^{2}}{\sqrt{2}})]),
\end{equation}
\begin{equation}
B(\xi ^{2})=3\xi ^{4}[\cosh (\xi ^{2})-\cos (\xi ^{2})-2\sinh (\frac{\xi ^{2}%
}{\sqrt{2}})\sin (\frac{\xi ^{2}}{\sqrt{2}})]+6C(\xi ^{2})+2D_{2}(\xi ^{2})
\end{equation}
and 
\begin{equation}
C(\xi ^{2})=\xi ^{2}[\sinh (\xi ^{2})-\sin (\xi ^{2})+\sqrt{2}(\sinh (\frac{%
\xi ^{2}}{\sqrt{2}})\cos (\frac{\xi ^{2}}{\sqrt{2}})-\sin (\frac{\xi ^{2}}{%
\sqrt{2}})\cosh (\frac{\xi ^{2}}{\sqrt{2}}))].
\end{equation}
We display in Fig. 4 the polar plot of $S_{\varphi ,N=4}^{(k=2)}$ at $|\xi
|=1.25$ as a function of $\varphi .$ In this case four directions of
squeezing (the shorter wings) are clearly seen. These four directions along
which squeezing is maximum are determined by $\varphi _{1}=\pi /8,$ $\varphi
_{2}=3\pi /8,$ $\varphi _{3}=5\pi /8$ and $\varphi _{4}=7\pi /8.$ For $N=8,$ 
$12,$ $16,$ $...$ simultaneous squeezing along the four above-mentioned
directions may also result with appropriate values of $\xi $ but the
squeezing is much less pronounced compared to the case with $N=4.$

\section{CASE $f\neq 1$}

In general $f$ is a nonlinear function of the boson number of a field. The
concrete form of $f$ depends on the physical system to be considered. For
example, for the so-called ``f-oscillator" \cite{r19,r19a} whose frequency 
is not a constant but depends on the number of the quanta of a field, 
$f$ can be given through $q$-deformed algebra \cite{q1,q2} as $f=\sqrt{\sinh
(n\lambda )/(n\sinh \lambda )}$ with $\lambda $ a real parametr or, for the so-called 
photon-added coherent states \cite{pacs} one has $f=1-m/(1+n)$ 
with the parameter $m$ being a positive integer \cite{r20}, etc. In
this section, we deal with a realistic situation associated with the phonon
field of the center-of-mass vibrational motion of a trapped ion. A laser-driven
scheme to produce the fan-state in this context has been presented in \cite
{ad}. In this specific physical system the function $f$ and the quantity $\xi $
are determined by
\begin{equation}
f=\frac{(n-2k)!L_{n-2k}^{2k}(\eta ^{2})}{n!L_{n-2k}^{0}(\eta ^{2})},
\label{f}
\end{equation}

\begin{equation}
\xi ^{2k}=-\frac{e^{i\phi }\Omega _{0}}{(i\eta )^{2k}\Omega _{1}}  \label{xi}
\end{equation}
where $L_{l}^{m}(x)$ is the $l$-th generalized polynomial in $x$ for
parameter $m$, $\eta $ is the Lamb-Dicke parameter, $\phi =\phi _{1}-\phi
_{0}$ with $\phi _{0}(\phi _{1})$ the phase of the driving laser which is
resonant with (detuned to the $2k$-th red sideband of) the electronic
transition of the trapped ion, and $\Omega _{0,1}$ are the Rabi frequencies.
As followed from Eqs. (\ref{f}) and (\ref{xi}), in this realistic system the function $f$ as well as
the magnitude of $\xi $ can be controlled by adjusting the parameters of the
driving lasers (through $\phi _{0,1}$, $\Omega _{0,1}$) 
and/or the trapping potential (through $\eta$). This provides an experimental means to tailor a
fan-state on demand. As for the question ``which power $N$ might give rise to
squeezing for a given $k",$ it is not affected by the concrete form of the
function $f$ since in fact this question depends merely on the property (\ref
{J}) of the $J_{k}(n)$ as argued in the preceding section. This means that
independent of $f$ the $N$-th power amplitude squeezing is possible for $%
N=2k,$ $4k,$ $6k,....$ However, squeezing actually arises only in certain
ranges of $\xi ^{2}$ and $\eta ^{2}.$ For each chosen value of $\eta$ there exists 
a critical $\xi_c(\eta)$ such that squeezing takes place when $0<|\xi|\leq|\xi_c(\eta)|$. 
This is illustrated in Figure 5 which plots $S_{\varphi,N=2}^{(k=1)}$ 
for $\varphi =\pi /4$ and $\eta ^{2}=0.05$ as a function of $%
\xi ^{2}$ : $\xi_c(\eta=\sqrt{0.05})=1.0099$. 
The number of squeezing directions is determined in the same way
as for $f\equiv 1$, i.e. squeezing if it exists is observed simultaneously
and equally maximal at $2k$ directions. It is worthy to emphasize that the number
of simultaneous squeezing directions is dictated only by $k$ but not by $N,
$ an interesting fact that can be justified from the symmetry property of the fan-state. 

The figure of merit of the case $f\neq 1$ in comparison with $f\equiv 1$ is that we can manage the
squeezing directions by controlling the physical parameters of the system
under consideration. Namely, we could control $\eta $ and $\xi $ so that 
$\langle a^{2N}\rangle _{k}>\max\{\langle a^{N}\rangle _{k}^2,
\langle a^{\mathbf{\dagger }N}a^{N}\rangle _{k}\}$
to have $2k$ squeezing directions determined by

\begin{equation}
\varphi _{j}=\frac{(2j+1)\pi }{4k}\hspace{1cm}\text{with}\hspace{0.1cm}%
j=0,1,...,2k-1.  \label{phi1}
\end{equation}
Or, if needed, we could chose $\eta $ and $\xi $ so that 
$\langle a^{2N}\rangle _{k}<\min\{\langle a^{N}\rangle _{k}^2,\,
2\langle a^{N}\rangle _{k}^2-\langle a^{\mathbf{\dagger }N}a^{N}\rangle _{k}\}$
to make squeezing along the other directions determined by

\begin{equation}
\varphi _{j}=\frac{\pi j}{2k}\hspace{1cm}\text{with}\hspace{0.1cm}%
j=0,1,...,2k-1.  \label{phi2}
\end{equation}
Transparently, the squeezing directions determined by Eqs. (\ref{phi1}) and (\ref{phi2}) 
are ``rotated" by $\pi/(4k)$ relative to each other. In other words, by adjusting 
the system driving parameters we are able to interchange the direction of squeezing and 
stretching if we like.  

\section{CONCLUSION}

We have dealt with the Hillery-type $N$-power amplitude squeezing in the
fan-state $|\xi ;2k,f\rangle _{F},$ Eq. (\ref{fan}), for both $f\equiv 1$
and $f\neq 1.$ Independent of $f$ we find out that for a given $k$ squeezing
may occur within a certain range of $\xi$ for powers $N$ that are a multiple of $2k.$ 
The remarkable feature 
in the fan-state is that whenever squeezing exists it arises simultaneously and equally 
in $2k$ directions, as opposed to conventional states in which squeezing can be 
observed only along one direction. While the number of squeezing directions
is fixed to $2k$ for a given $k,$ these directions themselves can be rotated 
by adjusting the physical system parameters when $f\neq 1$ as demonstrated 
in this work in the ion
trap context. Though it can be hoped, it seems premature to see that the multidirectional character 
of squeezing might find an actual application in practice in general or in quantum information 
processing in particular. For example, 
it is known that for quantum continuous variables teleportation \cite{t1,t2}  the 
necessary resource, i.e. the two-mode squeezed entangled state, is produced by superimposing on a 
beamsplitter two single-mode squeezed states whose squeezing directions must be perpendicular to 
each other. Yet, as at present it remains unknown what happens and what are the benefits if one 
or both of the superimposed states would exhibit squeezing in more than one direction. 
Such kinds of work are exciting and do deserve further efforts.

\begin{acknowledgments}
M.D.T. thanks the Abdus Salam ICTP for the financial support and hospitality at Trieste. 
B.A.N. is founded by KIAS R\&D Grant No. 03-0149-002.
\end{acknowledgments}

\newpage
\begin{figure}[t!]
\caption{$S=S^{k=1}_{\varphi,N=2}$ versus $|\xi|$ and $\varphi$ showing 
two coexistent directions of squeezing.} \label{fig1}
\end{figure}

\newpage
\begin{figure}[t!]
\caption{Polar plot of $S^{(k=1)}_{\varphi,N=2}$ for $|\xi|=0.8.$ The shorter 
wings along the directions $\varphi=\pi/4$ and $\varphi=3\pi/4$ are associated with 
squeezing, while stretching (longer wings) occurs along the directions $\varphi=0$ 
and $\varphi=\pi/2$.} \label{fig2}
\end{figure}

\newpage
\begin{figure}[t!]
\caption{The solid curve is the variance $\left< \left( \Delta Q_N(\varphi)\right)^2\right>_k$  
as a function of $\varphi$ for the same parameters as in Fig. 2, i.e. $k=1,$ $N=2$ and $|\xi|=0.8$. 
The dashed circle (of radius $\left< F_N\right>_k/4$) represents the corresponding coherent state case. } 
\label{fig3}
\end{figure}

\newpage
\begin{figure}[t!]
\caption{Polar plot of $S^{(k=2)}_{\varphi,N=4}$ for $|\xi|=1.25.$ Squeezing occuring simultaneously along 
the four directions (shorter wings) $\varphi=\pi/8$, $3\pi/8$, $5\pi/8$ and $7\pi/8$ are visual.}\label{fig4}
\end{figure}

\newpage
\begin{figure}[t!]
\caption{$S=S^{k=2}_{\varphi,N=4}$ versus $\xi^2$ for $\varphi=\pi/4$ 
and $\eta^2=0.05$ in the ion trap context.} \label{fig5}
\end{figure}

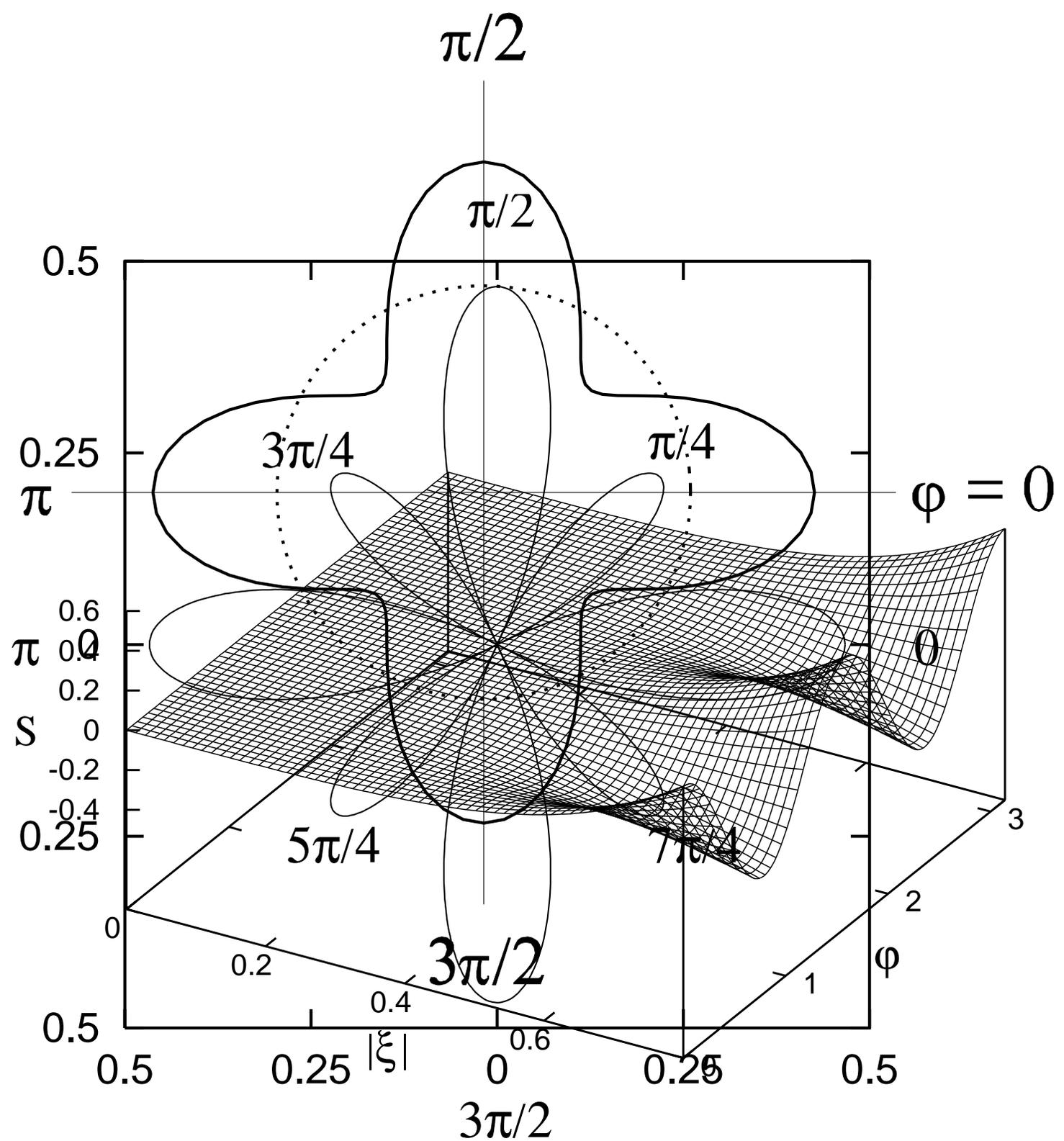
\end{document}